\documentclass[12pt,preprint]{aastex}
\usepackage{xcolor}

\shorttitle{Small-scale magnetic elements in Solar Cycle 23}
\shortauthors{Jin et al.}

\begin{document}

\title{The Sun\textsf{'}s small-scale magnetic elements in Solar
Cycle 23}

\author{C. L. Jin, J. X. Wang, and Q. Song}
\affil{Key Laboratory of Solar Activity of Chinese Academy of
 Sciences\\
 National Astronomical Observatories, Chinese Academy of
 Sciences, Beijing 100012, China;
 cljin@nao.cas.cn; wangjx@nao.cas.cn}

\and

\author{H. Zhao}
\affil{National Tsing Hua University, Hsinchu, Taiwan;
berserker0715@hotmail.com}

\begin{abstract}

With the unique database from Michelson Doppler Imager aboard the
Solar and Heliospheric Observatory in an interval embodying solar
cycle 23, the cyclic behavior of solar small-scale magnetic elements
is studied. More than 13 million small-scale magnetic elements are
selected, and the following results are unclosed. (1) The
quiet regions dominated the Sun\textsf{'}s magnetic flux for about 8
years in the 12.25 year duration of Cycle 23. They contributed (0.94
-- 1.44) $\times 10^{23}$ Mx flux to the Sun from the solar minimum
to maximum. The monthly average magnetic flux of the quiet regions
is 1.12 times that of active regions in the cycle. (2) The ratio of
quiet region flux to that of the total Sun equally characterizes the
course of a solar cycle. The 6-month running-average flux ratio of
quiet region had been larger than 90.0\% for 28 continuous months
from July 2007 to October 2009, which characterizes very well the
grand solar minima of Cycles 23-24. (3) From the small to large end
of the flux spectrum, the variations of numbers and total flux of
the network elements show no-correlation, anti-correlation, and
correlation with sunspots, respectively. The anti-correlated
elements, covering the flux of (2.9 - 32.0)$\times 10^{18}$ Mx,
occupies 77.2\% of total element number and 37.4\% of quiet Sun
flux. These results provide insight into reason for anti-correlated variations of small-scale magnetic activity during the solar cycle.

\end{abstract}

\keywords{Sun: magnetic fields --- Sun: photosphere --- Sun: sunspots}

\section{Introduction}

No any other astrophysical process but solar cycle leaves
massive footprints on human's living environment. This eleven-year
cycle was discovered by a Germany pharmacist Schwabe (1843) from the
number changes of solar sunspots. A primary understanding on the
solar cycle has been established based on the theories and
simulations of a mean-field magnetohydrodynamic (MHD) dynamo
(Charbonneau 2005). However, new observations are continuously
challenging our understanding by the myriad of new and seemingly
conflicting observations. A more severe challenge comes from
observations of small-scale magnetic elements (see de Wijn et al.
2009). Therefore, it is still a difficult task to
explore the physics of solar cycle.

Since the 1960s it has been observed that small-scale magnetic
fields outside of sunspots are everywhere on the Sun (Sheeley 1966,
1967; Harvey 1971). The stronger magnetic elements at the boundaries
of supergranulation cells are network elements, while the smaller
and weaker elements within the supergranulation cells are
intra-network (IN) elements (Livingston \& Harvey 1975; Smithson
1975). Similar to emerging flux regions (EFRs) in sunspot groups (or
active regions) (Zirin 1972), small-scale emerging bipoles named
ephemeral (active) regions (ERs) were described by Harvey
and Martin (1973). They account for the formation of network
elements in addition to the debris from decaying sunspots. It was
noticed that the flux emerging rate in ER exceeds that in sunspots
by two orders of magnitude (Zirin 1987). Moreover, flux generation
rate of IN elements exceeds that of ERs by another two orders of
magnitude. Further smaller magnetic fibrils are believed to be
mostly unresolved by present telescopes, yet their aggregation is
the dominant mechanism by which IN and network elements
appear (Lamb et al. 2008). A substantial amount of solar magnetic
flux is probably still hidden (Trujillo et al. 2004).

As soon as the small-scale magnetic elements were identified, great
efforts have been made to understand how they change during a solar
cycle and if they are correlated with sunspots. Diverse observations
are reported, igniting discussions and debates in the
literature. The observations are made either directly from the
magnetic measurements, or indirectly from proxies of small-scale
magnetic flux, e.g., the G-band and CaII K bright points and coronal
X-ray bright points. Key revelations are listed below.

\begin{enumerate}

\item[(1)] No cyclic variations: CaII K emission in solar quiet
regions (White \& Livingston 1981); modern X-ray bright points observations (Sattarov et al. 2002; Hara \&
Nakakubo 2003); magnetic flux of networks (Labonte \& Howard 1982);
flux spectrum and total flux of network elements with flux $\leq
2.0\times 10^{19}$ Mx (Hagenaar et al. 2003); Stokes $\frac{Q}{I}$
profile (Trujillo et al. 2004).

\item[(2)] Anti-correlation of small-scale fields with sunspot
cycle: number of network bright points in very quiet regions (Muller
\& Roudier 1984, 1994); HeI 10830\ \AA \ dark points in the higher
chromosphere (Harvey 1985); early X-ray bright points observations (Davis et
al. 1977; Davis 1983; Golub et al. 1979); Weak
changes of emergence frequency of ERs with flux less than
(3-5)$\times 10^{19}$ Mx (Hagenaar et al. 2003).

\item[(3)] Correlation with sunspot cycle: more ERs appeared during
active solar condition (Harvey \& Harvey 1974; Harvey 1989); the
number (or magnetic flux) of network structures (Foukal et al. 1991;
Meunier 2003); flux distribution and total flux of network
concentrations with flux $(2.0-3.3)\times 10^{19}$ Mx (Hagenaar et
al. 2003).

\end{enumerate}

The observations listed above are related to some
fundamental, but not yet resolved questions in solar physics: the
origin, dynamics and active role in Sun's global processes of solar
small-scale magnetic elements, as well as the controlling physics of
solar activity cycle. However, discrepancy among different authors
is not yet understood, implying problems either in the
observations or on the physics used to interpret the observations. A
few aspects make things even more complicated.

First, for the observations of the proxies of small-scale
magnetic elements, the connections between the magnetic elements and
their proxies are not well quantified, and the underlined physics is
not known exactly. There seems to be not a one-to-one correspondence
between network elements and network bright points (Zhao et al.
2009). In other words, the widely-adopted paradigm of ``magnetic
bright points" is still questionable. Moreover, the early revelation
about the magnetic properties of coronal X-ray bright points (Golub
et al. 1977), needs to be revisited and updated with
state-of-the-art observations.

Secondly, quite many reports listed above went back to
early solar observations, which makes us difficult to
evaluate the quality of the observations. We are confused
by rather poor resolution, calibration and consistency in
sensitivity in early magnetic measurements. As an example,
the early Mont. Wilson magnetograph observations (Labonte \& Howard
1982) were with a resolution of $\geq$ 12.5-17.5 arcsec, and the
calibration was not consistent time to time. We are simply not able
to say anything confidently about their conclusions. Additionally, early X-ray bright point measurements, which suffered from low cadence, purported to show a decrease in the number of X-ray bright points with the solar cycle. More recent higher cadence observations have called into question whether this effect is real. It reminds to be seen whether other observations of variation with the solar cycle also need to be reinterpreted. New observations with careful and thorough data reduction and interpretation are crucially required.

Thirdly, even for recent observations, sometimes, the
different algorithm and logic in data analysis make us hard to judge
the results too. An interesting example comes from the analysis of
full-disk magnetograms of the Michelson Doppler Imager aboard the
Solar and Heliospheric Observatory (MDI/SOHO) (Scherrer et al.
1995). By adopting the different detection algorithm and approaches,
Meunier (2003) revealed correlation of the network element number
(or flux) with sunspots; in contrast, Hagenaar et al. (2003)
declared some weak anti-correlated emergence rate of ERs and an
independence of the total absolute flux for smaller network
concentrations. This discrepancy should be clarified with
new analysis.

To clarify the problem and to close the debates are an essential
task in understanding the solar cycle phenomena. Fortunately, now
MDI/SOHO is providing a unique database - the full-disk magnetograms
over more than 13 years, covering the complete 23rd Solar Cycle. The
13.5 year 5-min average full disk magnetograms are used in the
current study. However, the poor temporal resolution makes the
identity of ERs questionable and the sensitivity of the full-disk
magnetograms rules out the possibility to resolve the IN elements.
Therefore, what we have identified in this study is basically the
network magnetic elements.

In this paper, we aim at learning the
cyclic variations of quiet Sun's magnetic flux and small-scale magnetic elements.  To use the full-disc MDI magnetograms with the
temporal coverage of entire Cycle 23 comes from an awareness of the
intermittency of solar cyclic behavior in both the temporal and
spatial domains. By the intermittency to select the magnetograms of a short interval, e.g., 10-30 hours, in a month for each year, at the `supposed' different cycle phases. would not guarantee a grape of the key characteristics of a solar cycle. From our understanding, to choose the database that cover the entire cycle 23 is of overwhelming importance. The database for the current study is unique in the sense that it is the only space-borne magnetic measurements of the full Sun, for which the consistency in sensitivity and resolution persisted for a cycle-long interval. As we are interested in the global behavior of small-scale magnetic elements, sampling network elements in a cycle-long temporal domain and in all different flux ranges (or strengths) are more important than selecting a few high cadence sequences interruptedly. Moreover, the magnetic elements with different flux (or size) may have different origins and characteristics, therefore we group all the network magnetic elements into different categories in accordance with their magnetic flux.

In section 2, we describe the observations, the technique of
calibration, the evaluation of noise level of the magnetograms, the
separation of active regions and the quiet Sun, and the selection of
network elements. In section 3, we present the results of cyclic
behavior of quiet region magnetic flux and small-scale magnetic
elements. In section 4, we make the comparison with previous studies, and
consider the possibilities on how to understand the anti-correlated network magnetic elements with sunspots. In section 5, we draw the conclusions.

\section{Observations and methods}

The MDI instrument aboard SOHO spacecraft provides the full-disk
magnetogram with a pixel size of 2''. In order to obtain a low noise
level, only those 5-min average magnetograms are selected in the
study. We extract one observed full-disk magnetogram per day, and
thusly we totally select 3764 magnetograms from 1996 September to
2010 February, which include the complete 23rd solar cycle. In order
to further reduce the noise level, we apply a boxcar smoothing
function to each magnetogram by a width of 6''$\times$6''. There are
two groups of authors who first pointed out the under-estimation of
magnetic flux by earlier MDI full-disk magnetogram calibration
(Berger et al. 2003; Wang et al. 2003). All the magnetograms used in
this study are that retrieved after recalibration of December 2008.
For a better understanding about the cyclic behavior of solar minima
of Cycles 22 and 23, we extend the MDI data base by adding Kitt Peak
full-disk magnetograms from August 1996 back to January 1994.
The data merging is made based on a least-square fitting of
the mean flux density of Kitt Peak magnetograms to that of MDI
magnetograms for the common interval of 1996.

We estimate the noise level of these smoothed 5-min average
magnetograms according to the method described by Hagenaar (2001)
and Hagenaar et al. (2003). Based on these magnetograms, we analyze
their histograms of magnetic flux density. The core of the
distribution function is fitted by a Gaussian function $
F(x)=F_{max}exp(-x^2/2\sigma^2)$, where the width $\sigma$ of the Gaussian function, about 6 Mx/cm$^{2}$, is
defined as the noise level.

We assume that the observed line of sight magnetic flux density is a
projection of the intrinsic flux density normal to the solar
surface, so the magnetic flux density for each pixel is
corrected(see Hagenaar 2001 and Hagenaar et al. 2003) as
$B_{cal}=B_{obs}(\alpha)/\cos(\alpha)$. The angle $\alpha$ of each pixel is defined by
$\sin(\alpha)=\sqrt{x^{2}+y^{2}}/R$
%\begin{subequations}
%\begin{equation}
%   B_{cal}=B_{obs}(\alpha) /cos(\alpha)
%\end{equation}
%\begin{equation}
%   sin(\alpha)=\sqrt{x^{2}+y^{2}}/R.
%\end{equation}
%\end{subequations}
Where x and Y are the pixel position referring to the disk center,
at which x and y is equal to 0, and the R is the solar disk radius.
After the correlation, the magnetogram shows the magnetic flux
density normal to solar surface.

After the angle is greater 60 degrees, there are less and less
magnetic signals due to the lower magnetic sensitivity and spatial
resolution of MDI/SOHO magnetograms, and the magnetic noise
level would increase according to the magnetic correction
1/cos($\alpha$). Therefore, we only analyze these pixels with angle
$\alpha$ less than 60 degree, i.e., the region included by the black
circle in the left panels of Fig. 1. The flux density of the pixel
with 60$^{o}$ $\leq$ $\alpha$ $\leq$ 90$^{o}$ is set to zero.

For each smoothed and corrected full-disk magnetogram of MDI/SOHO,
we apply a magnetic flux density of 15 Mx/cm$^{2}$ as a threshold to
define the active regions and their surroundings, and then create a
mask for each magnetogram. These masks include many small clusters
and isolated pixels, so only the islands with area larger 9$\times$9
pixels are defined as the active regions (Hagenaar et al. 2003).
Considering the active regions close to the edge of 60 degree, in
order to avoid missing them in the automatic procedure, we
always search the active regions in the solar disk with
angle $\alpha$ less than 70 degree first, as that
shown in the left panels of Fig. 1. Thusly, the islands with area
less than 81 pixels within 60 degree disk are still defined as the
active regions if they have more than 81 pixels searched within 70
degree disk.

Two magnetograms within the 70$\deg$ from disk center at
approximately the solar maximum and minimum phases, respectively,
are displayed in the left panels of Fig. 1. On these retrieved
magnetograms the selected ARs are masked by red curves. The
criterion of selecting ARs appears to work well from a visually
examination for the given cases. In the right panels two
selected sub-windows of the magnetograms are shown with contours
outlining the network elements which are selected by a
procedure of automatic feature selection. The yellow and green
contours outline the selected network elements that are belong to the components of correlated
and anti-correlated with sunspots in the solar cycle, respectively
(see Section 3.2).

\section{Results}
\subsection{Cyclic variations of magnetic flux of solar quiet regions}

In order to compare the cyclic variations of magnetic flux of active
regions with that of quiet regions, we calculate their magnetic
flux, respectively, which is shown in the left panel of Fig. 2. At
the same time, the area ratio of quiet regions is also computed,
which is shown by purple \textsf{`}+\textsf{'} symbols in the right
panel of Fig. 2. It is found that the quiet Sun contributed
$(0.94-1.44) \times 10^{23}$ Mx flux from approximately the solar
minimum to maximum in Cycle 23. The fractional area of quiet regions
always exceeds 80\% in the entire solar cycle 23, and decreased from
the cycle minimum to maximum by a factor of 1.2, although their
total flux increased by a factor of 1.53; as a comparison, the
active region flux increased by several orders of magnitude. The
measurements confirm the global behavior of the quiet Sun fields
(see Meunier 2003 and Hagenaar et al. 2003). During the 12.25 years
of Cycle 23, from October 1996 to December 2008 (see
http://www.ips.gov.au), the quiet Sun dominated the Sun\textsf{'}s
magnetic flux for 7.92 years. The monthly average magnetic flux of
quiet Sun is 1.12 times that of active regions. The magnetic fields
on the quiet Sun, indeed, are a fundamental component of the
Sun\textsf{'}s activity cycle which maintains the Sun\textsf{'}s
magnetic energy and Poynting flux at a certain level.

It is interesting to notice that the ratio of the quiet
Sun\textsf{'}s magnetic flux to solar total flux (referring to as
the flux occupation by the quiet Sun) equally characterizes the
course of a solar cycle, like sunspots. The occupation of 6-month
running-average magnetic flux by the quiet Sun is shown by purple
cross symbols in the right panel of Fig. 2. The active region flux
shown in the left panel answers for the variation of sunspot cycle
very well. However, for the quiet regions, the maximum occupation of
magnetic flux marks the minima of solar cycles. For instance, in our
data set, the maximum flux occupation of quiet Sun, which was
96.0\%, first happened in October of 1996 at the beginning of Cycle
23. The later maxima happened from July 2008 to August 2009. In
December of 2008, the beginning of Cycle 24, the maximum occupation
reached 99.3\%. The 6-month running-average fractional flux of quiet
Sun had been larger than 90.0\% for 28 continuous months (from July
2007 to October 2009), which characterizes the grand solar minima of
Cycles 23-24. Staying at such a low activity level there were 25
months, for which the total AR flux was less than $10^{22}$ Mx.
However, during the minima of Cycles 22-23 for only intermittent 7
months, i.e., from December 1995 to April 1996 and from December
1996 to January 1997, we had witnessed the fraction larger than
90.0\%. The distinction between two solar minima are so severe,
which can be seen very clearly in Fig. 2.

\subsection{Cyclic variations of network magnetic elements}

After excluding the active regions, we apply the magnetic noise,
i.e., 6 Mx/cm$^{2}$ as a threshold to create a mask for each quiet
magnetogram, and define these magnetic concentrations with more than
10 pixels in size as network magnetic elements (Hagenaar et al.
2003). More than 13 million network elements have been identified
for the interval from September 1996 to February 2010. The
probability distribution function (PDF) of these magnetic elements
in the studied interval is shown in Fig. 3 as the average flux
distribution. From the figure, it can be found that the distribution
of magnetic flux of network magnetic elements mainly concentrate at
the flux of 10$^{19}$ Mx. This peak distribution is consistent with that found for multiple MDI full-disk datasets by parnel et al. (2009) (see their Fig. 5) 

For an exclusive examination, we divide all the magnetic elements
into 96 sub-groups according to the flux per element. In this way, a
statistical sample is created, covering the range of magnetic flux
per element from the smallest observable network flux of $1.5\times
10^{18}$ Mx for the current data set to an upper limit of
$3.8\times 10^{20}$ Mx. The monthly variation of magnetic elements
for each sub-group is calculated and examined in term of number
density and absolute total flux in the interval from October 1996 to
February 2010, embodying the entire Cycle 23. The influence of the
area changes of the quiet Sun on both quantities has been removed.
There are 0.3\% of network elements (or clusters) with flux larger
than the upper limit, which were fragments of decayed sunspots and
not included in the sample. By following Hagenaar et al. (2003) tiny
flux pieces with less than 10 pixels are not considered in the
study. As a whole the total flux of these tiny flux pieces showed a
small variation in the scope of $(3.5-4.0) \times 10^{22}$ Mx during
the cycle.

The correlation coefficients between the cyclic variation of numbers
of network elements and sunspots are calculated for each sub-group
of network elements and shown in Fig. 4. They are the linear
Pearson correlation coefficients of two vectors for each sub-group
elements. Denote the element number in sub-group $i$ as $N_i$ and
the sunspot number $N_s$, then the correlation coefficient between
$N_i$ and $N_s$ will be

\begin{equation}
\rho(N_i,N_s)=Covariance(N_i,N_s)/(Variance{N_i}\times
Variance{N_s})^\frac{1}{2}
\end{equation}

The confidence level about the correlation can be found in
some basic statistics handbook by taking account of how big was of
the sample. As the sample size for each sub-group elements is 162,
which is quite large. If the coefficient is higher than 0.256,
then the failure probability of the linear correlation would already
be $<0.001$. 

From the small to the large flux spectrum, there appears a
remarkable 3-fold correlation scheme between the network elements
and the sunspots: basically no-correlation, anti-correlation and
correlation. This behavior is held for both the element number and
total flux. Either the anti-correlation or the correlation has been
observed at very high confidence level. The majority of the
correlations show a failure probability $\leq 0.001$. Between the
anti-correlation and correlation, there is a narrow range of
magnetic flux per element of (3.2 - 4.3)$\times 10^{19}$ Mx. Network
elements falling in this flux range show a transition from
anti-correlation to correlation with sunspot cycle (see the narrow
shaded column in the middle of Fig. 4.)

The dependence of the correlation coefficient on the element flux
hints the possibility that network elements at different segments of
the flux spectrum may present different physical origins and
different cyclic behavior accordingly. For an detailed examination
of cyclic behavior of network elements, we group all the network
elements into 4 categories which show, respectively, no-correlation,
anti-correlation, transition from anti-correlation to correlation,
and correlation with sunspot cycle. For each category, its flux
range, percentage in number and in total flux, as well as the
correlation coefficient with sunspots are listed in Table 1. We,
then, discriminate the cyclic variation of magnetic elements in
accordance to the flux range listed in the table. The detailed
cyclic variations of each category network elements are shown in
Fig. 5.

Approximate 77.2\% of the magnetic elements, covering the flux range
of (2.9-32.0)$\times 10^{18}$ Mx show anti-correlation with the
sunspot cycle. This anti-correlated component contributes 37.4\% of
network flux during Cycle 23. Transition from anti-correlation to
correlation takes place between (3.20 - 4.27) $\times 10^{19}$ Mx.
The correlated component elements have magnetic flux larger than
4.27$\times 10^{19}$ Mx. They occupy approximately 15.7\% in number
but 53.5\% of total flux of network elements. In the flux range of
(1.5-2.9)$\times 10^{18}$ Mx, network elements show randomly
independent variation with the sunspot cycle. From this data set,
they occupy less than 0.6\% of network elements and have neglectable
total flux. With the poor sensitivity in flux measurements
at the smallest end of the flux spectrum, it could not be excluded
that the non-correlation component manifested some random noises in
flux measurement. More serious efforts with higher resolution and
sensitivity data are necessary to clarify the cyclic behavior of
smallest observable magnetic elements.

The number changes of the network elements in the flux range of (2.9
- 32.0)$\times 10^{18}$ Mx show obviously anti-phase correlation
with sunspot cycle, so do the changes of their total unsigned flux. However, the
cyclic minimum of this anti-correlation component is not exactly
coincided with the reversed profile of the maximum of the
sunspot cycle, implying complexity in causing the anti-correlation.
Meanwhile, the flux changes of the magnetic elements with flux
larger than 4.3$\times 10^{19}$ Mx show remarkable in-phase
correlation with sunspot cycle. The same is true that the
profiles of the maximum of network elements and that of sunspots are
not corresponding one another exactly. There is a 5-7 month delay
of their cyclic maximum related to that of the sunspot cycle. This seems to be related to the characteristics dispersal time of active region fields.

To further explore the cyclic variation of magnetic elements, we
obtain the PDFs of yearly network magnetic elements according to the
magnetic flux, and compute the differential PDFs, i.e., the
difference between the yearly PDFs and average PDF (see
Fig.3). Here, the differential probability distribution function
is abbreviated as DPDF. We plot the DPDFs, and show the variation
for magnetic flux spectrum from 1996 to 2010 in Fig. 6. From the
figure, we confirm the 3-fold scenario of cyclic variations of
network elements.

From the solar minimum to solar maximum (see the first
column of the figure), the distribution of magnetic elements
in the flux range about (3-30)$\times 10^{18}$ Mx gradually
decrease, which shows the anti-correlation variation with the
sunspot cycle; while the distribution of magnetic elements of flux
larger than about 4$\times 10^{19}$ shows the correlation variation
with the sunspot cycle and reaches the peak in the years 2000, 2001
and 2002. Furthermore, the distribution of magnetic elements with
flux of $\sim$ 3$\times 10^{19}$ and less than 3$\times 10^{18}$ Mx
shows almost no variation. The distribution of magnetic elements
correlated with the sunspot cycle reaches the smallest values in the
years 2007, 2008 and 2009, which are the solar minima of cycle
23-24; while the distribution of magnetic elements anti-correlated
with the sunspot cycle shows outstanding peak during this long
interval (see the third column of the figure). The
distribution characterizes the long duration of the solar minima of
Cycles 23-24. It is noticed that the distributions in some
of the ascending and declining phases (see that in 1998 and 2005)
are, more or less, represent the average distribution of small-scale
magnetic elements shown in Fig.3.

\section{Discussion}

With the unique space-borne observations which comprised a
complete solar cycle, we have revealed a 3-fold correlation scheme
of the Sun's small-scale magnetic elements with sunspot cycle, and
identified an anti-correlation component of network elements that
dominates the element population. Before coming up with conclusion
and discussion on the physics, a comparison with previous studies
that adopted the similar approaches and with that same space-borne
MDI observations (see Section 2) is necessary.

Hagenaar et al.(2003) selected high cadence magnetograms of
6 time-sequences, each of which covered 10-30 hours in a month from
1996 to 2000. These authors found that the component of network
elements with flux $\geq 30\times 10^{18}$ Mx varied in phase with
the sunspot cycle. The magnetogram calibration
they adopted had under-estimated the flux density by a factor of
about 1.6 (Bergers et al. 2003; Wang et al. 2003). With the renewed
calibration, this component would consist of magnetic elements with
flux $\geq 4.8\times 10^{19}$. This is a component in our analysis
that changes in phase with sunspots. Meunier (2003) chose strong
magnetic elements with threshold flux density of 25 G and 40 G,
respectively, and found naturally a correlation of number and flux
of network elements with sunspots.

When the element flux $\leq 20\times 10^{18}$ Mx (i.e.,
32$\times 10^{18}$ Mx in renewed calibration), Hagenaar et al.(2003)
declared that both the flux spectrum of quiet network elements and
the total flux changed a little with the cycle phase. These authors
used the interrupted data in the 6 years of the ascending cycle
phase, they would not be able to guarantee a grasp of the real trend
of the cyclic modulation. We tested their results by using the same
6-month 5-m magnetograms, and found a weak change, but anti-phased
with the cycle phase, in both numbers and flux for network elements
in this flux range. In fact, Hagenaar et al.(2003) reported that the
number density of network concentrations on the quiet Sun decreased
by less than 20\% from 1997 to 2000, consistent with our approaches.
They also suggested an even anti-correlated changes in flux
emergence rate in this low flux range. The revelation of a
remarkable anti-correlation component of network elements with a
broad flux range from several times of $10^{18}$ Mx to 3 times of
$10^{19} Mx$ is likely to be the true nature of small-scale solar
magnetism and inspiring new considerations of the Sun's magnetism.

Exploration of the magnetic nature of the Sun's small-scale
activity went back to earlier solar studies. A few pioneer studies
stand still as reliable references in solar physics. Mehltretter
(1974) identified that the network bright points represented
magnetic flux concentration with field strength of 1000-2000 G, each
of them had a mean flux of 4.7$\times 10^{17}$ Mx. Later, the work
was extended by Muller \& Roudier (1984, 1994). They deduced an
average flux of 2.5$\times 10^{17}$ Mx for an network bright points.
The flux range suggested by these authors for the network bright
points changing anti-phased with sunspot cycle, is out of reach by
current data base. Muller \& Roudier (1984) also identified the
correlation between the network bright points in the photosphere and
coronal XBPs. For the latter, Golub et al.(1977) carefully studied
their magnetic properties, and found the average total flux
associated with a typical XBP was 2.0$\times 10^{19}$ Mx. The
magnetic measurements were obtained at Kitt Peak with a fine scan of
2.5 arcsec resolution element and $\sim$2G noise level. They are
reasonably reliable to quantify the magnetic flux of an XBP. This
typical flux, even with some uncertainty, e.g., 50\% or larger, is
still falling in the flux range of the anti-correlated component of
network elements discovered by this study. We tentatively suggest
that the anti-correlated component of magnetic elements are
responsible for the small-scale activity, e.g., the coronal X-ray
bright points. Updated efforts to quantify the magnetic properties
of the so-called magnetic bright points are crucial to final resolve
the long-lasting puzzle of the anti-phase behavior of the Sun's
small-scale activity in a solar cycle.

Observationally, small-scale network elements come from
several sources: fragmentation of active regions, flux emergence in
the form of ephemeral regions, coalescence of intranetwork flux, and
products of dynamic interaction among different sources of magnetic
flux. The 3-fold relationship between network elements and sunspot
cycle has immediate implication on the Sun\textsf{'}s magnetism. As
demonstrated by state-of-the-art simulations (see
V\"{o}gler \& Sch\"{u}ssler 2007), the magnetic elements at the
smallest end of the flux spectrum, either resolved or un-resolved,
manifest a local turbulent dynamo which operates in the
near-photosphere and is independent to the sunspot cycle. On the
other hand, at the larger flux end, the magnetic elements are likely
to be the debris of decayed sunspots. They follow, of course, the
solar cycle.

The key issue here is how to understand the majority of magnetic
elements which are anti-correlated with sunspots in the solar cycle.
They are not likely the debris of decayed sunspots, but probably
created by turbulent local dynamo action that, however, is globally
affected or controlled by the sunspot field from the mean-field MHD
dynamo. A few possibilities now are being considered.

First, during the more active times of the Sun, the smaller magnetic
elements created by the turbulent dynamo have more opportunity to
encounter sunspots and their fragments. The same-polarity
encountering results in a merging of those elements to the flux
related to sunspots. Whereas, the opposite polarity encountering
causes flux cancelations with the net results of lost smaller
elements and a diffusion of sunspot flux. What accompanied the
sunspot flux diffusion is the reduced smaller elements with the
turbulent origin. This accounts for the anti-correlated magnetic
component possibly. By this kind of interaction magnetic flux from
turbulent dynamo actively takes part in the operation of the solar
cycle, helping with more efficient magnetic diffusion. To
quantify this mechanism, studies of dynamic interaction between
small-scale magnetic elements and active regions fields are
crucially required.

Secondly, it is also possible that at the solar maximum, the
stronger magnetic field from sunspots tends to suppress the
Sun\textsf{'}s global convection in some measure. As a result, the
local dynamo has been abated somehow, and the network elements
created by turbulence are reduced in number and total flux. This
seems to suggest that the turbulent dynamo is, in fact, global
but not local. Unfortunately, so far there have been no
definite observations about the changes in the global solar
convection during the sunspot cycle.

Another possibility is that the anti-correlated component represents
the recycling of parts of the previously diffused or submerged
magnetic flux from the mean-field dynamo (Parker 1987). The
diffusion of magnetic flux from sunspots to the deep convection zone
requires 5-7 years (Jiang et al. 2007). Parts of the diffused or
submerged flux serves as the seed field for the globally turbulent
dynamo. Its production is naturally out of phase with sunspots in
the solar cycle, and brings up the magnetic elements that
anti-phased with sunspots.

In a recent literature, Thomas and Weiss (2008) proposed a picture
of the solar Dynamo on three scales (one large and two small),
which, according to the above authors, were only loosely coupled to
each other. It is not clear if some unknown interplay of different
scale dynamos may result in the complicated behavior of the Sun's
small-scale fields. If we adopt the common vision that the smaller
magnetic elements are created by a local turbulent dynamo,
then the local turbulent dynamo on a certain scale must
have closely correlated to the global mean-field dynamo. The global
dynamo either provides seed flux or modifies the condition for this
\textsf{`}global\textsf{'} turbulent dynamo. At the smallest end,
the dynamo is likely to be more \textsf{`}local\textsf{'}. The
turbulent dynamo, either global or purely local, brings a tremendous
amount of turbulent flux to the Sun that continuously interacts with
the products of the mean-field dynamo. The interaction seems to not
only help with the operation of the global dynamo, but also power
the ceaseless small-scale magnetic activity and maintain the
Sun\textsf{'}s Poynting flux to Earth and interplanetary space.

\section{Conclusions}

With the unique database from MDI/SOHO in the interval from
September 1996 to February 2010, which embodies the entire Solar
Cycle 23, we analyze the cyclic variations of quiet Sun's magnetic
flux and Sun\textsf{'}s small-scale magnetic elements.

The quiet regions contributed $(0.94-1.44) \times 10^{23}$ Mx flux
from approximately the solar minimum to maximum in Cycle 23. The
fractional area of quiet regions decreased from the cycle minimum to
maximum by a factor of 1.2, but their total flux increased by a
factor of 1.53. The quiet regions dominate Sun's magnetic flux over
60\% duration of the cycle. Furthermore, the ratio of the quiet
region magnetic flux to the Sun's total flux can be used to describe
the course of solar cycle, just as sunspots. The maximum flux
occupation of quiet regions marks the minima of solar cycle. The
flux occupation on the quiet Sun had been larger than 90\% for 28
continuous months from July 2007 to October 2009, which seems to
equally characterize the grand minima of Cycles 23 and 24.

With increasing magnetic flux per element the number and total flux
of the Sun's small-scale magnetic elements follow no-correlation,
anti-correlation and correlation changes with sunspots. The
anti-correlated component, covering the flux range of (2.9 -
32.0)$\times 10^{18}$ Mx, occupies 77.2\% of total elements and
37.4\% of flux on the quiet Sun. However, the stronger
magnetic elements with flux larger than 4.3$\times 10^{19}$ Mx
dominate the quiet Sun magnetic flux and follow closely the sunspot
cycle.

The definitively identified anti-correlated component of the
small-scale magnetic elements seems to offer an
interpretation on the puzzling observations of anti-correlation
variation of many types of small-scale activity with the solar
cycle, e.g., the network bright points, HeI 10830 \AA \ dark points
and coronal X-ray bright points. 

It is speculated that the
anti-correlated small-scale magnetic elements are products of some
local turbulent dynamo or dynamos that is modulated to be
anti-phased with the global mean-field dynamo.

\acknowledgments

The authors are grateful to Dean-Yi Chou, Sara Martin and
Jie Jiang for their valuable suggestions and discussions. We appreciate the instructive advice and valuable suggestions of the anonymous referee, by which the paper has been significantly improved.
The work is supported by the National Natural Science Foundation of
China (10873020, 11003024, 40974112, 40731056, 10973019, 40890161, 10921303,
11025315), and the National Basic Research Program of China
(G2011CB811403).

\begin{table}
\caption{Cyclic variation of the NT elements with different flux
range\label{tbl}}
\begin{tabular}{llllr}
\hline
 Category & Flux (in Mx) & Number ratio & Flux (ratio) & Cor. \\
\hline

No-correlation & (1.5--2.9)$\times 10^{18}$& 0.58\% & 6.48$\times 10^{21}$ (0.05\%)& -0.04\\
Anti-correlation&(2.9--32.0)$\times 10^{18}$& 77.19\% &  4.72$\times 10^{24}$ (37.40\%)& -0.45\\
Transition& (3.20--4.27)$\times 10^{19}$& 6.59\% &  1.15$\times 10^{24}$  (9.08\%) & -0.03\\
correlation&(4.27--38.01)$\times 10^{19}$& 15.65\% &  6.74$\times 10^{24}$ (53.46\%) & 0.82\\
\hline
\end{tabular}
\end{table}

\begin{figure}
\includegraphics[scale=0.8]{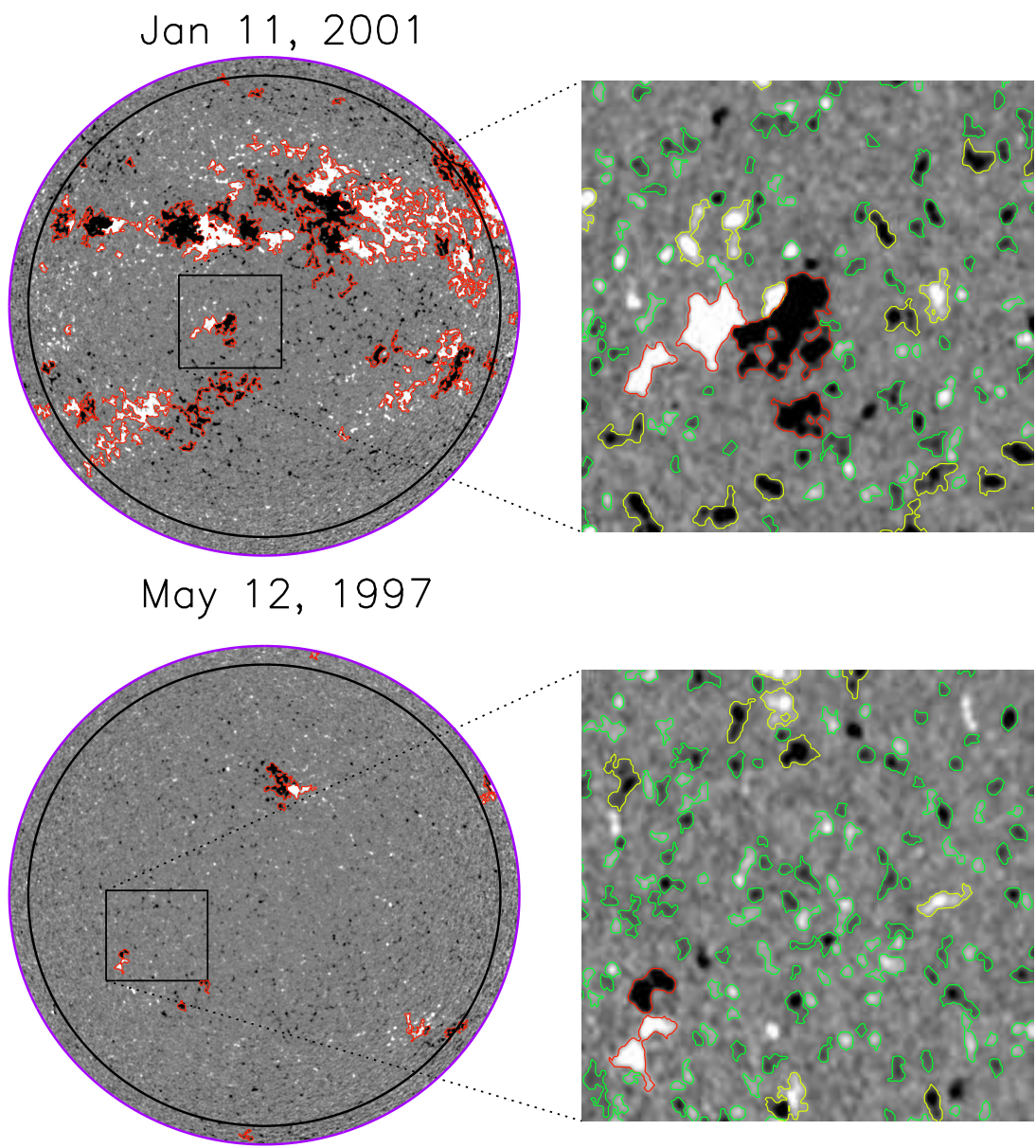}
\caption{Left panels: Two retrieved MDI 5-minute full-disk
magnetograms within 70 $\deg$ from disk center, at approximately the
solar maximum and minimum, respectively. Using the threshold on 15
Mxcm$^{-2}$ to define the edge of active region, the islands with
area larger 9$\times$9 pixels, i.e., the regions contoured by red
line, is defined as active regions. The purple circle displays the
location $\alpha$=70 $\deg$, and the black circle displays the location
$\alpha$=60 $\deg$. The gray scale saturates at $\pm$50 Mxcm$^{-2}$.
Right panels: enlarged images for the windows framed in the MDI
magnetograms, on which network elements falling in the flux ranges
of (2.9-32.0)$\times$ 10$^{18}$ Mx and (4.3-38.0)$\times$ 10$^{19}$ Mx
are outlines by green and yellow curves, respectively (see Section
3.2).}
\end{figure}

\begin{figure}
\includegraphics[scale=0.8]{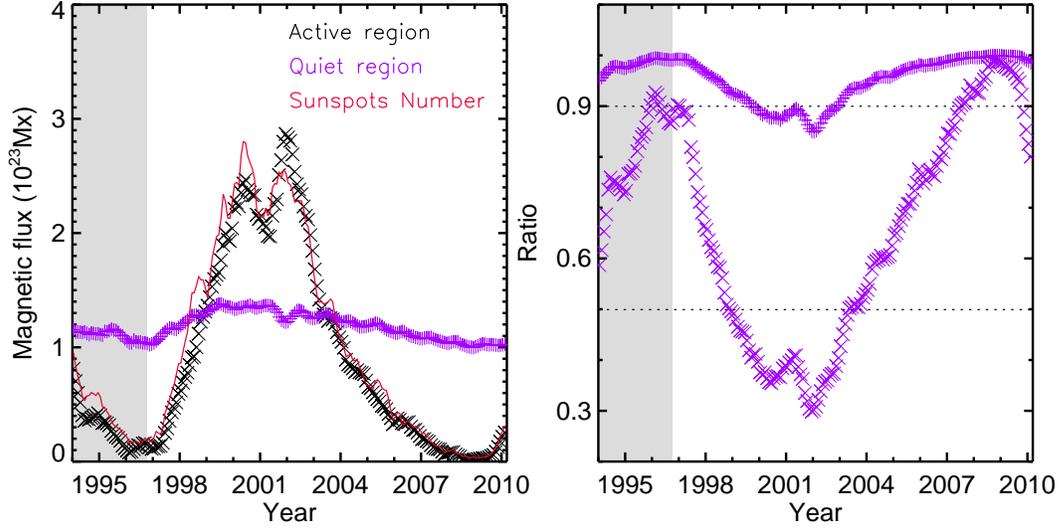}
\caption{The left panel is the flux variations of ARs
(cross symbols in black) and quiet Sun
(\textsf{`}+\textsf{'} symbols in purple) in an interval including
the entire 23rd Solar Cycle. The red curve represents the sunspot number changes in the cycle. The shaded columns are the statistical
results based on the Kitt Peak full-disk magnetograms. The magnetic
flux for quiet regions rises from 0.94$\times 10^{23}$ Mx in
December 1995 to 1.44$\times 10^{23}$ in May 2002, increases by a
factor of 1.53. The fractional quiet Sun area is shown by purple
\textsf{`}+\textsf{'} symbols, in the right panel.
It decreases by a factor of 1.2 from the solar minima to maximum.
The ratio of quiet Sun flux to the total Sun's flux, the flux
occupation of the quiet Sun, is shown by purple
cross symbols in the right panel.
The quiet Sun flux has dominated the Sun's magnetic flux for 7.92
years in the 12.5 year Cycle 23.}
\end{figure}

\begin{figure}
\includegraphics[scale=0.8]{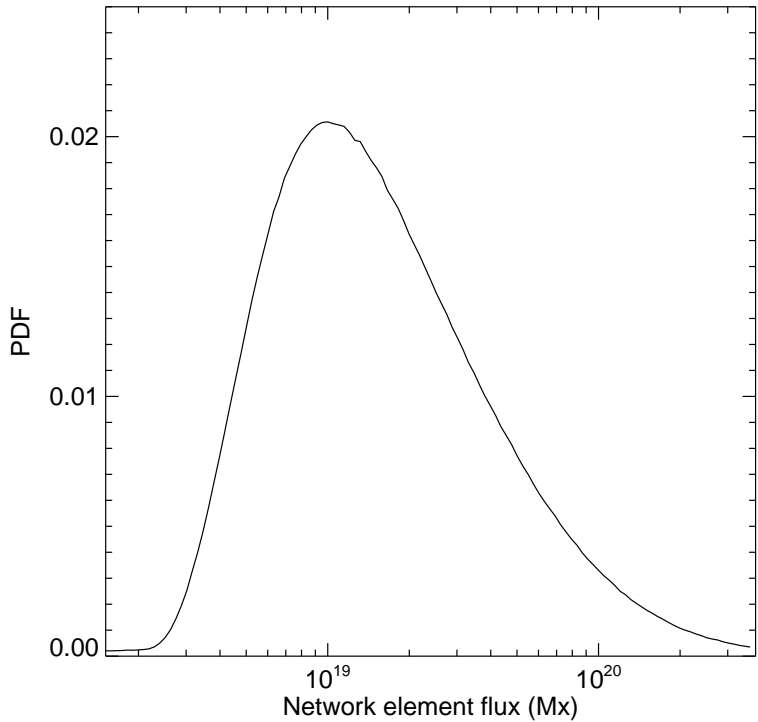}
\caption{The probability distribution function of element magnetic
flux for all the selected network elements during the interval from
September 1996 to February 2010, i.e., average PDF of the quiet Sun's small-scale magnetic elements. The peak distribution at $10^{19}$ Mx is consistent with that found for multiple MDI full-disk datasets by Parnell et al. (2009). }
\end{figure}

\begin{figure}
\includegraphics[scale=0.8]{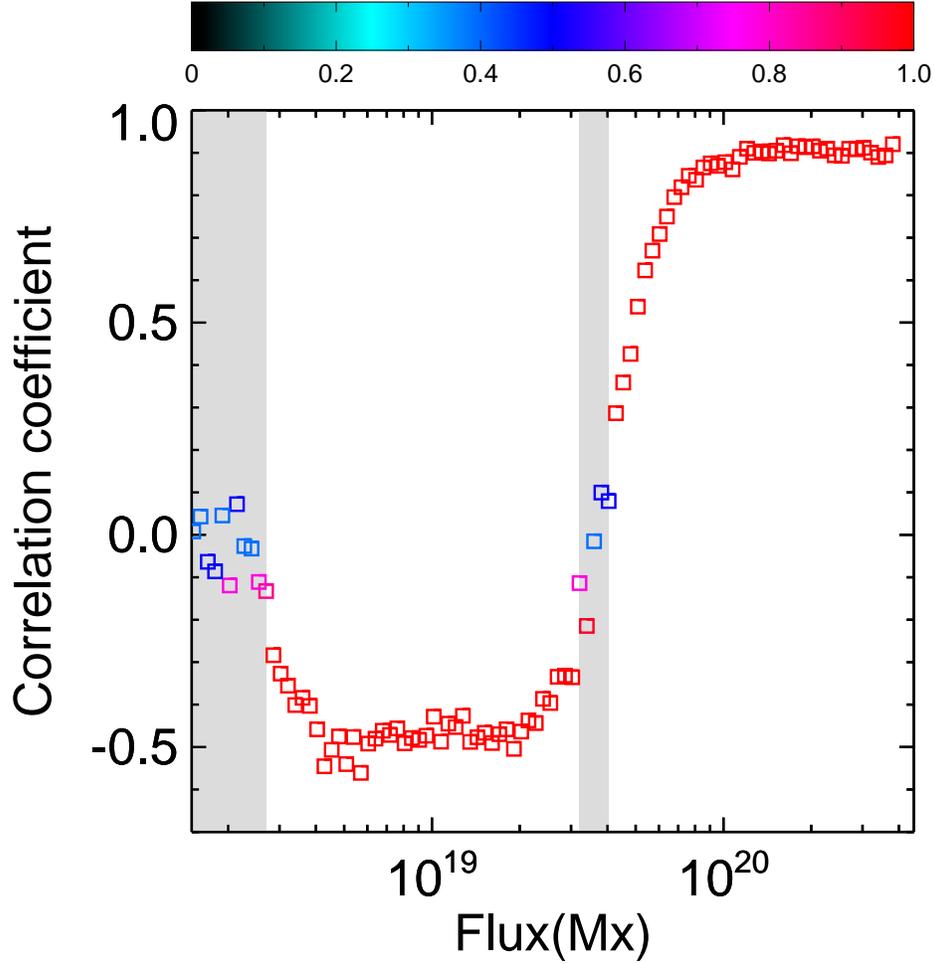}
\caption{Correlation coefficients between the sunspot number and
network element number of each of the 96 sub-group elements which
are reconstructed according to the flux per element. There appears a
3-fold correlation scheme between the network elements and the
sunspot cycle: basically no-correlation, anti-correlation and
correlation. At the low end of flux spectrum, there are very small
correction coefficients. With the increasing flux per element, the
correlation coefficients reach approximately to -0.58, then they
become positive and reach as high as 0.92 after a very narrow
transition in the flux range of (3.20-4.27)$\times 10^{19}$ Mx. The color bar represents the confidence level.}
\end{figure}

\begin{figure}
\includegraphics[scale=0.8]{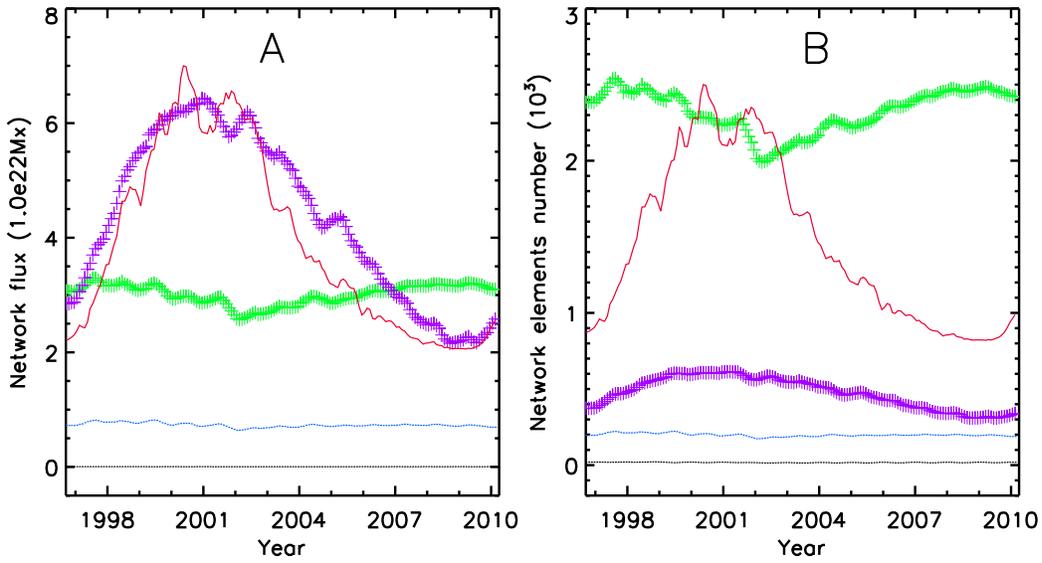}
\caption{Cyclic variations of network element number (right panel)
and flux (left panel) of 4 categories of network elements shown in
Table 1, which represents the 3-fold correlation scheme of network
elements with the sunspot cycle. The green `+' is
referring anti-correlation component elements, while the purple
`+' is for the in-phase correlation component
elements. Black and blue dotted lines are elements which have no
correlation or shown transition from anti-correlation to correlation
with the solar cycle.}
\end{figure}

\begin{figure}
\includegraphics[scale=0.9]{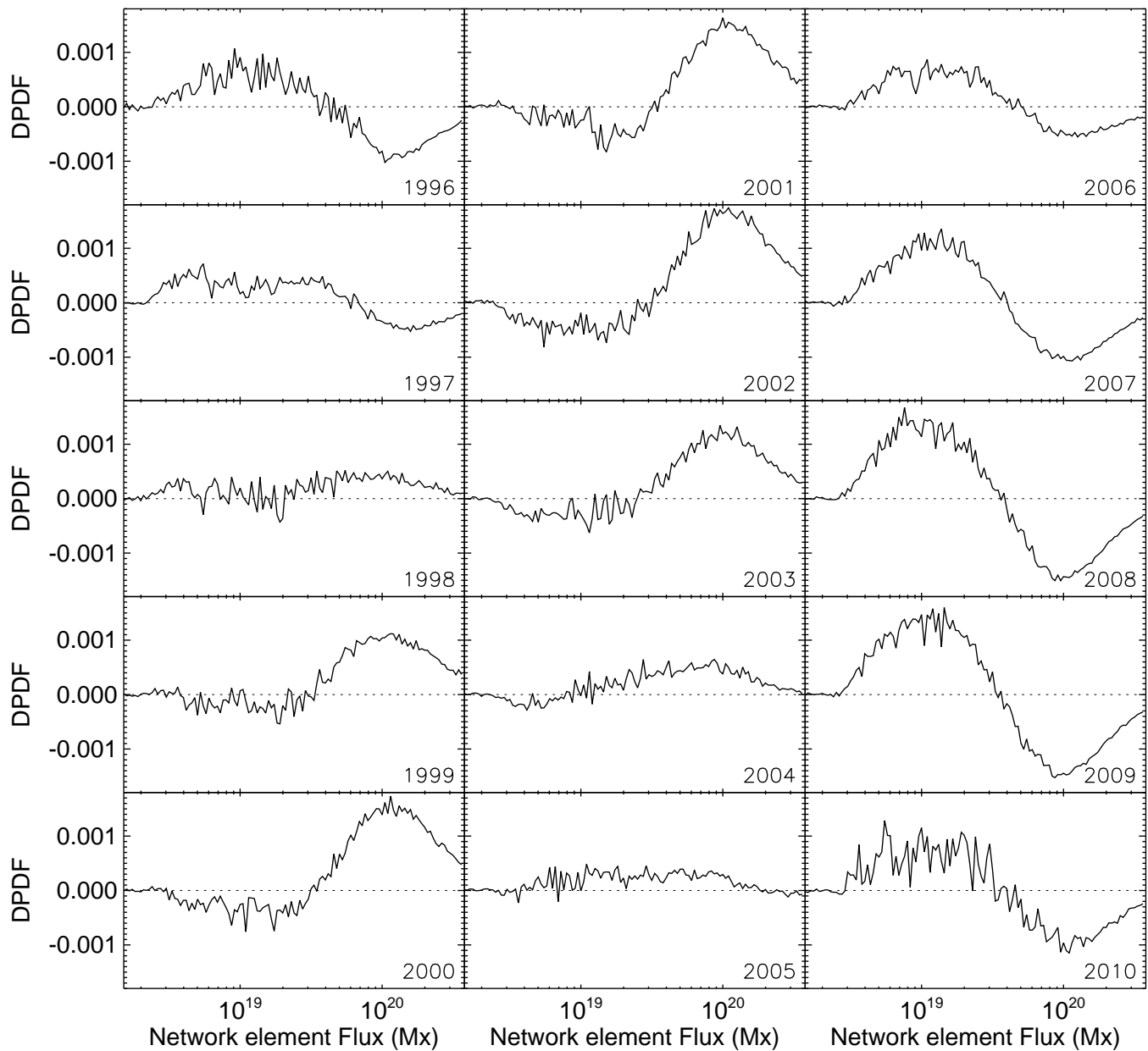}
\caption{The differential probability distribution function (DPDF), i.e.,
the difference between the PDFs of
yearly network magnetic elements and average PDF shown in Fig.3.}
\end{figure}

\end{document}